\documentstyle [aps,prl,epsf,multicol]{revtex} 

\newcommand{\half}   {\frac{1}{2}}
\newcommand{\asix}   {\frac{a^2}{6}}
\newcommand{\rr   }   { {\bf r} }

\newcommand{\dr  }   { {\rm d}\rr\ }

\newcommand{\e   }   { {\rm e} }

\begin{document}
\draft

\title {\bf Scaling Laws of Polyelectrolyte Adsorption}

\author {Itamar Borukhov and David Andelman$^*$}
\address{School of Physics and Astronomy,
         Raymond and Beverly Sackler
         Faculty of Exact Sciences,
         Tel-Aviv University,
         Ramat-Aviv 69978, Tel-Aviv, Israel}

\author {Henri Orland}
\address{Service de Physique Th\'eorique,
         CEA, CE-Saclay,
         91191 Gif-sur-Yvette Cedex, France }

\date{\today}

\maketitle

\begin{abstract}
  Adsorption of charged polymers (polyelectrolytes) 
from a semi--dilute solution to a charged surface
 is investigated theoretically.
We obtain simple scaling laws for (i) the amount of 
polymer $\Gamma$ adsorbed to the surface  
and (ii) the width $D$ of the adsorbed layer,
as  function of the fractional charge per monomer $p$ 
and the salt concentration $c_b$. 
For strongly charged polyelectrolytes 
($p \alt 1$) in a low--salt solution, 
both $\Gamma$ and $D$ scale as $p^{-1/2}$.
In salt--rich solutions $D\sim c_b^{1/2}/p$
 whereas the scaling behavior of $\Gamma$ 
depends on the strength of the polymer charge.
For weak polyelectrolytes ($p\ll 1$) we find that 
 $\Gamma\sim p/c_b^{1/2}$ while 
 for strong polyelectrolytes $\Gamma\sim c_b^{1/2}/p$.
Our results are in good agreement with 
adsorption experiments and with numerical solutions
of mean--field equations.
\end{abstract}



\pagestyle{plain}
\section{Introduction}

Polyelectrolytes (charged polymers) 
are widely used in 
industrial applications.
For example, many colloidal suspensions
can be stabilized by the adsorption of
polyelectrolytes.
In many experiments, the total amount 
of polymer adsorbed on a surface 
(the polymer {\em surface excess}) 
is measured as a function of 
the bulk polymer concentration, pH and/or  
ionic strength of the bulk solution
\cite{Peyser,Kawaguchi,Meadows,Denoyel,Blaakmeer,vandeSteegExp,Shubin,Hoogeveen}.
(For reviews see, e.g.,  refs.~\cite{CS-1988,CS-chapter,FleerBook,Norde}).
More recently, spectroscopy \cite{Meadows} and
ellipsometry \cite{Shubin} 
have been used to measure
the width 
of the adsorbed polyelectrolyte layer.
Other techniques  such as  
neutron scattering
can be employed to measure the entire profile
of the adsorbed layer 
\cite{Auvray,Guiselin}.

   The theoretical treatment of polyelectrolytes in solution 
is not very well established because of the delicate 
interplay between the chain connectivity and 
the long range nature of electrostatic interactions
\cite{Oosawa,deGennesBook,Odijk,Dobrynin}.
   In many studies adsorption of polyelectrolytes is treated as an 
extension of neutral polymer theories.
In these approaches the polymer 
concentration profile is  determined by  minimizing 
the overall free energy.
   
One approach is a discrete {\em multi--Stern layer} model 
\cite{vanderSchee,Papenhuijzen,Evers,vandeSteegSim,Linse}, where
the system is placed on a lattice whose sites can be occupied
by a monomer, a solvent molecule or a small ion. 
The electrostatic potential is determined
self--consistently together with the concentration profiles
of the polymer and the small ions.
   Another approach treats the electrostatic potential 
and the polyelectrolyte concentration as continuous functions
\cite{Muthukumar,Varoqui91,Varoqui93,epl,long_bulk}. 
These quantities are obtained from two coupled differential equations 
derived from the total free energy of the system.

In the present work we focus on the adsorption behavior 
of polyelectrolytes near a single charged surface held at a constant
potential.
 Simple scaling expressions are presented and compared 
to concentration profiles that we obtain from exact numerical
solutions, 
and to experiments measuring the amount of polymer 
adsorbed on the surface.
In Sec.~I the adsorption problem is treated numerically.
We then present in Sec.~II simple scaling arguments describing 
the adsorption characteristics and
in Sec.~III we compare our scaling results to experiments.
Finally, we present our conclusions and some future prospects.

\section{Numerical Profiles}


   Consider a semi--dilute solution of polyelectrolytes in 
good solvent in contact with a charged surface (Fig.~1). 
In addition to the polymer chains and their counterions, 
the solution contains 
small ions (salt) assumed hereafter to be  monovalent. 
The system is coupled to a bulk  reservoir containing
polyelectrolyte chains and salt.
   In the present work we assume that
the charge density on the polymer chains
is continuous and uniformly distributed 
along the chains.
This assumption is valid as long 
as the electrostatic potential is not too high, $|\beta e\psi|<1$,
where $1/\beta=k_B T$ is the thermal energy,
 $e$ is the electron charge and 
 $\psi$ is the electrostatic potential.
Further treatments of the polymer charge distribution
 (annealed and quenched models)
can be found in refs. \cite{epl,long_bulk}. 

   Within mean--field approximation, the free energy of the 
system can be expressed in terms of the local electrostatic potential 
 $\psi(\rr)$, the local monomer concentration $\rho_m(\rr)$
and the local concentration of positive and negative ions
 $c^{\pm}(\rr)$.
It is convenient to introduce the polymer 
order parameter $\phi(\rr)$ where $\rho_m(\rr)=|\phi(\rr)|^2$.
The excess free energy with respect to the bulk $F$ is
then \cite{Varoqui91,Varoqui93,epl,long_bulk} 
\begin{eqnarray}
   F &=& \int \dr \left\{ f_{pol}(\rr) + f_{ions}(\rr) 
                        + f_{el}(\rr) \right\}
   \label{F}                   
\end{eqnarray}
  The polymer contribution is
\begin{eqnarray}
   f_{pol}(\rr)  &=& k_BT \left[  \asix|\nabla\phi|^2
                 + \half v (\phi^4-\phi_b^4) \right]
                 - \mu_p (\phi^2-\phi_b^2)
        \label{fpol}
\end{eqnarray}
where the first term is 
the polymer elastic energy,
$a$ being the effective monomer size.
The second term is the excluded volume contribution where 
 $v\sim a^3$. The last term couples the system to the reservoir, 
where $\mu_p$ is the chemical potential of the polymers and
 $\rho_m(\infty)=\phi_b^2$ is  the bulk monomer concentration.

   The entropic contribution of the small (monovalent) ions is
\begin{eqnarray}
   f_{ions}(\rr) &=& \sum_{i=\pm} k_BT
                   \left[ c^i\ln c^i - c^i
                        - c^i_b\ln c^i_b + c^i_b \right]
                 - \mu^i (c^i-c^i_b)
        \label{fent}
\end{eqnarray} 
   where $c^i(\rr)$, $c^i_b$ and $\mu^i$ are, respectively, 
the local concentration, the bulk concentration and 
the chemical potential of the $i=\pm$ ions.

Finally, the electrostatic contributions is 
\begin{eqnarray}
   f_{el}(\rr)  &=& p e\phi^2\psi +e c^+\psi -e c^-\psi
                 - \frac{\varepsilon}{8\pi}|\nabla \psi|^2
        \label{fel}
\end{eqnarray} 
  The first three terms are the 
electrostatic energies of the monomers, 
the positive ions and the negative ions, respectively, 
 $p$ is the fractional charge carried by one monomer.
  The last term is the self energy of the electric field where
 $\varepsilon$ is the dielectric constant of the solution.
Note that the electrostatic contribution, eq.~\ref{fel},
is equivalent to the 
well known result: 
$F_{el}=(\varepsilon/8\pi) \int \dr |\nabla \psi|^2$ 
plus surface terms. 
This can be seen
by substituting  the Poisson--Boltzmann equation 
(as obtained below) into eq.~\ref{fel} and then integrating 
by parts.

Minimization of the free energy with respect to $c^\pm$, $\phi$
and $\psi$ yields a Boltzmann  distribution
for the density of the small ions,
 $c^\pm(\rr)=c^\pm_b \exp(\mp\beta e\psi)$, and two
coupled
differential equations for $\phi$ and $\psi$:
\begin{eqnarray}
   \nabla^2\psi(\rr)
   &=& \frac{8\pi e}{\varepsilon}c_b \sinh(\beta e\psi)
    -  \frac{4\pi e}{\varepsilon}
       \left( p\phi^2 - p\phi_b^2 \e^{\beta e\psi} \right)
   \label{PBs} \\
   \asix \nabla^2\phi(\rr) 
   &=& v(\phi^3-\phi_b^2\phi) + p\phi\beta e\psi 
   \label{SCFs}
\end{eqnarray}
   Equation~\ref{PBs} is a generalized 
Poisson--Boltzmann equation
including the free ions as well as the charged polymers.
The first term represents the salt contribution and
the second term is due to the 
charged monomers and their counter-ions.
 Equation~\ref{SCFs} is a generalization of the self--consistent 
field equation of neutral polymers \cite{deGennesBook}.
In the bulk, the above equations are satisfied by
setting $\psi\to 0$ and $\phi\to \phi_b$. 


   When a polyelectrolyte solution is in contact with a
 charged surface, 
the chains will  adsorb to (or deplete from)
the surface, depending on the nature of the monomer--surface 
interactions.
The large number of monomers on each polymer chain enhances 
these interactions.
For simplicity, we assume that the surface is ideal, {\it i.e.}, 
flat and homogeneous.  In this case
physical quantities depend only on 
the distance $x$ from the surface (see Fig.~1).
The surface imposes 
boundary conditions on the polymer order parameter $\phi(x)$ 
and electrostatic potential $\psi(x)$.
In thermodynamic equilibrium all charge carriers
in solution should exactly balance the surface charges 
(charge neutrality).
The self--consistent field equation, 
the Poisson--Boltzmann equation 
and the boundary conditions 
uniquely determine the polymer concentration profile
and the electrostatic potential. 
In most cases, these two  coupled non--linear equations
can only be solved  numerically. 

In the present work we have chosen the surface to be 
at a constant potential $\psi_s$, leading to the following 
electrostatic boundary condition \cite{elec_bc}

\begin{eqnarray}
  \left. \psi \right|_{x=0} = \psi_s
  \label{BCpsi0}
\end{eqnarray}
The boundary conditions for $\phi(x)$ depend on the nature of 
the short range interaction of the monomers and the surface.
For simplicity, we take a non--adsorbing surface
and require  that the monomer concentration will  vanish there:
\begin{eqnarray}
  \left. \phi \right|_{x=0} = 0
  \label{BCphi0}
\end{eqnarray}
 Far from the surface 
($x\rightarrow\infty$) 
both $\psi $ and $\phi $ reach their bulk values
and their derivatives vanish: 
 $ \psi'|_{x\rightarrow\infty} = 0$ and  
 $\phi'|_{x\rightarrow\infty} = 0$.

Figure~2 shows several adsorption profiles
obtained from numerical solutions of the mean--field equations 
(eqs.~\ref{PBs}, \ref{SCFs}) using  a minimal squares method.
  The polymer is positively charged and is attracted to  
the non-adsorbing surface
 held at a constant negative potential.
The aqueous solution contains a small amount of monovalent salt
($c_b=0.1$mM). 
The reduced concentration profile $c(x)/\phi_b^2$ 
is plotted as a function of the distance from the surface.
 Different curves correspond to different values of 
the reduced surface potential $y_s=\beta e\psi_s$, 
the charge fraction $p$  and
the monomer size $a$.
   Although the spatial variation of the  profiles
differs in detail, they all have a single 
peak characterized by an adsorption length. 
We use this feature in the next section to 
obtain simple analytical expressions characterizing the 
adsorption.

\section{Scaling Results}

  The difficulty in obtaining a simple picture of 
polyelectrolyte adsorption lies in the existence of 
several length scales in the problem: 
 (i) the Edwards correlation length $\xi=a/({v\phi_b^2})^{1/2}$,
 characterizing the concentration fluctuations of 
neutral polymer solutions;
(ii) the Debye-H\"uckel screening length 
$\kappa_s^{-1}={(8\pi l_B c_b)}^{-1/2}$ 
where $l_B=e^2/\varepsilon k_B T$
is the Bjerrum length equal to about 7$\AA$\ 
for aqueous solutions at room temperature.
Additional length scales  can be associated with  
electrostatic and/or short range  surface interactions.

  Motivated by the numerical results (Fig.~2), 
we assume that the balance between 
these interactions results in one dominant 
length scale $D$ characterizing the adsorption at the surface. 
Hence, we write the polymer order parameter profile in the form of
\begin{eqnarray}
   \phi(x)=\sqrt{c_m}h(x/D)
   \label{scaling_prof}
\end{eqnarray}
where $h(z)$ is a dimensionless function 
normalized to one at its maximum and
 $c_m$ sets the scale of polymer adsorption.
The free energy can be now expressed  in terms of $D$ and $c_m$
while the exact form of $h(z)$ affects only the
numerical prefactors. 

In principle, the adsorption length $D$ depends also on the 
ionic strength through $\kappa_s^{-1}$.
As discussed  below  the scaling assumption 
(eq.~\ref{scaling_prof}) is only valid 
as long as $\kappa_s^{-1}$ and $D$ are not of the same 
order of magnitude. Otherwise, $h$ 
should depend on both $\kappa_s x$ and $x/D$.
We concentrate now on two limiting regimes where 
eq.~\ref{scaling_prof} can be justified:
(i) the low--salt regime  $D\ll\kappa_s^{-1}$ and
(ii) the salt--rich regime  $D\gg\kappa_s^{-1}$.

\subsection{Low--Salt Regime; $D\ll\kappa_s^{-1}$}

In the low--salt regime the effect of the small ions can be 
neglected and the free energy, eqs.~\ref{F}-\ref{fel}, is approximated by
(see also ref.~\cite{Varoqui93})
\begin{eqnarray}
  \beta F = A_1{a^2\over 6D}c_m - A_2 p|y_s|c_m D
          + 4\pi B_1 l_Bp^2 c_m^2D^3
          + \half B_2 v c_m^2 D
  \label{scalingF}
\end{eqnarray}
  The first term is the elastic energy 
characterizing the response of the polymer to concentration 
inhomogeneities. The second term accounts for the 
electrostatic attraction of the polymers to the charged surface.
 The third term represents the Coulomb repulsion 
between adsorbed monomers. Indeed, the interaction 
between two layers
with surface charge densities $\sigma=pe\phi^2(x)dx$ and 
 $\sigma'=pe\phi^2(x')dx'$ 
is  proportional to the distance $|x-x'|$ yielding
the $D^3$ dependence.  
The last term represents the excluded volume repulsion 
between adsorbed monomers, where we assume that 
the monomer concentration near the surface is much 
larger than the bulk concentration $c_m\gg\phi_b^2$ .
The coefficients $A_1,A_2,B_1$ and $B_2$ are numerical
prefactors, which depend on the exact shape of 
the dimensionless function $h(z)$.
These coefficients can be explicitly calculated for a specific profile 
by integrating the Poisson equation without taking into account
the small ion contributions \cite{coefficient}.
For a linear profile, 
 $h(z)=z$ for $0\le z\le 1$ and $h(z)=0$ for $z> 1$, 
we get $A_1=1,A_2=1/3,B_1=1/14$
and $B_2=1/5$;
For a parabolic profile,
 $h(z)=4z(1-z)$ for $0\le z\le 1$ and $h(z)=0$ for $z> 1$, 
 we get
 $A_1=16/3,A_2=8/15,B_1\simeq 1/9$ and $B_2\simeq 2/5$.

   In the low--salt regime and for 
strong enough polyelectrolytes 
the electrostatic interactions 
are much stronger than the excluded volume ones.
Neglecting the latter interactions and minimizing the free energy 
with respect to $D$ and $c_m$ gives:
\begin{eqnarray}
  D^2 &=& {5A_1\over 6A_2} {a^2 \over p|y_s|} 
      \sim {1\over p|y_s|}
  \label{scalingD}
\end{eqnarray}
  and
\begin{eqnarray}
  c_m &=& {12A_2^2\over 25A_1B_1} {|y_s|^2 \over 4\pi l_B a^2}
      \sim |y_s|^2
  \label{scalingcm}
\end{eqnarray}
The above expressions are valid as long as 
(i) $D\ll\kappa_s^{-1}$
and (ii) 
the excluded volume term in eq.~\ref{scalingF} is negligible.
The former condition translates into
$c_b\ll p |y_s|/(8\pi l_B a^2)$.
For 
$|y_s|\simeq 1$, $a=5$$\AA$\ and $l_B=7$$\AA$\ 
this limits  the salt concentration to
$c_b/p \ll $ 0.4 M. 
The latter  condition on the 
magnitude of the excluded volume term
can be shown to be equivalent to
$p\gg v|y_s|/l_B a^2$. 
These
requirements are consistent with the data presented
in Fig. 2. 


We recall that the 
profiles presented in Fig.~2 were obtained from the 
numerical solution of eqs.~\ref{PBs} and \ref{SCFs}, 
including the 
effect of small ions and excluded volume.
 The scaling relations 
 are verified by plotting in 
 Fig.~3 the same sets of data as in Fig.~2
using rescaled variables as defined in 
eqs.~\ref{scalingD},\ref{scalingcm}. Namely,
the rescaled electrostatic potential
 $\psi(x)/\psi_s$ and polymer concentration 
 $c(x)/c_m\sim c(x)a^2/|y_s|^2$ are plotted
as functions of the rescaled  
distance $x/D\sim x p^{1/2}|y_s|^{1/2}/a$.
The different curves roughly collapse on the same curve.

  In many experiments the total amount of adsorbed polymer
per unit area $\Gamma$
is measured. Our scaling assumption yields
\begin{eqnarray}
  \Gamma = \int_0^\infty [c(x)-\phi_b^2] dx 
         \simeq D c_m 
         \simeq {|y_s|^{3/2} \over l_B ap^{1/2}}
         \sim {|y_s|^{3/2} \over p^{1/2}}
 \label{scalingGamma1}
\end{eqnarray}

   The adsorbed amount $\Gamma(p)$ in the low--salt regime 
is plotted in the inset of Fig.~4a.
 One of the important features
 is the decrease in $\Gamma$
with increasing charge fraction $p$.
This can be understood in the following way:
the  monomer--monomer Coulomb repulsion scales
as $(p c_m)^2$, and dominates over 
the adsorption energy  scaling only as
$p c_m$.

\subsection{Salt--Rich Regime; $D\gg\kappa_s^{-1}$}

  The opposite case occurs when  $D$ is
much larger than $\kappa_s^{-1}$.
 In this case the electrostatic
interactions are short ranged 
with a cut-off $\kappa_s^{-1}$ \cite{Varoqui93}.
The free energy then reads:
\begin{eqnarray}
  \beta F = A_1{a^2\over 6D}c_m - A_2 p|y_s|c_m\kappa_s^{-1} 
          + 4\pi B_1 l_Bp^2 \kappa_s^{-2}c_m^2 D
          + \half B_2 v c_m^2 D
  \label{scalingF2}
\end{eqnarray}
Note that $\kappa_s^{-1}$ enters in the 2nd and 3rd terms. 
The third term can be viewed as an 
additional electrostatic excluded volume
 with $v_{\rm el} \sim l_B (p/ \kappa_s)^2$. 

   Minimization of the free energy gives
\begin{eqnarray}
  D &=& {A_1\over 2A_2} {\kappa_s a^2\over p|y_s|}
     \sim {c_b^{1/2} \over p|y_s|}
  \label{scalingD2}
\end{eqnarray}
  and 
\begin{eqnarray}
  c_m &\sim& { p^2 |y_s|^2/(\kappa_s a)^2 \over 
                    B_1 p^2/c_b + B_2 v}
  \label{scalingcm2}
\end{eqnarray}
  yielding 
\begin{eqnarray}
  \Gamma &\sim& {p |y_s|c_b^{-1/2} \over 
                    B_1 p^2/c_b + B_2 v}
  \label{scalingGamma2}
\end{eqnarray}

 The adsorption behavior is depicted in Figs.~4 and 5.
Our results are in agreement with numerical solutions of
discrete lattice models  (the multi--Stern layer theory)
\cite{CS-1988,CS-chapter,FleerBook,vanderSchee,Papenhuijzen,Evers,vandeSteegSim,Linse}.
In Fig.~4  $\Gamma$ is plotted 
as function of $p$ (Fig.~4a) 
and pH (Fig.~4b) for three different salt concentrations.
  The behavior as seen on Fig.~4b represents annealed
 polyelectrolytes 
where the nominal charge fraction is controlled 
by the pH of the solution through
\begin{eqnarray}
  p = {10^{{\rm pH-pK}_0} \over 1 + 10^{{\rm pH-pK}_0} }
  \label{pHpK}
\end{eqnarray}
where ${\rm pK}_0 = -\log_{10}{\rm K}_0$ and ${\rm K}_0$ is the 
apparent dissociation constant. 

Another interesting observation which can be deduced from
eq.~\ref{scalingGamma2} 
is that $\Gamma$ is only a function  of 
$p/\sqrt{c_b}$. Indeed, as can be seen in Fig.~4,
$c_b$  only affects the position 
of the peak and not its height.

The effect of salt concentration is shown in Fig.~5, where 
  $\Gamma$ is plotted in 5a as function of 
the salt concentration $c_b$ for two charge fractions 
 $p=0.01$ and $0.25$. In Fig.~5b, $D$ is plotted for the same
range of $c_b$ and for $p=0.1$.
The solid curves are obtained within
the salt--rich regime 
(eq.~\ref{scalingGamma2}).

The extrapolation of the salt--rich expression 
eq.~\ref{scalingGamma2} towards
low values of $c_b$ does not give the correct
low--salt limit, because the basic assumptions
of the salt--rich regime are no longer valid. Instead,
a simple interpolation
\cite{interpolation} 
between the low--salt  and salt--rich  regimes
is used in the same figure (dashed curves). It
 demonstrates a plausible behavior
of $\Gamma$ and $D$ for intermediate 
salt concentrations where our above 
scaling expressions are not valid.
At low salt concentrations,
$\Gamma$  is almost independent of $c_b$ and saturates to
the low--salt value (left hand side of Fig.~5a).
 At high  salt concentrations, the 
salt--rich result is recovered (right hand side of Fig.~5a). 
For weak polyelectrolytes (e.g., $p=0.01$ in Fig.~5a), 
addition of salt weakens the surface attraction. Consequently,
$\Gamma$ is a decreasing function of $c_b$
in the whole $c_b$ range. 
For strong polyelectrolytes, (e.g., $p=0.25$ in
Fig.~5a), $\Gamma$ is an increasing function of $c_b$
at low salt concentrations and a decreasing function
at high salt concentrations. As a result, there is a maximum
in $\Gamma$ at some intermediate value of $c_b$.

 From Figs.~4, 5a and   eq.~\ref{scalingGamma2}, 
it is clear
that  the salt--rich regime can be divided into two sub regimes 
according to the polyelectrolyte charge.
At
low charge fractions (sub-regime SR I), 
$p\ll p^*=(c_b v)^{1/2}$, 
the  excluded volume term 
dominates the denominator of eq.~\ref{scalingGamma2} and
\begin{eqnarray}
\Gamma &\sim& {p|y_s| \over c_b^{1/2}} 
  \label{scalingSRI}
\end{eqnarray}
whereas at high $p$ (sub-regime SR II),
 $p\gg p^*$,
the monomer--monomer electrostatic repulsion dominates
and $\Gamma$ decreases with $p$
and increases with $c_b$:
\begin{eqnarray}
 \Gamma &\sim& {c_b^{1/2} |y_s| \over p}  
  \label{scalingSRII}
\end{eqnarray}
The various regimes with their crossover lines are shown 
schematically in Fig.~6.


\section{Comparison to Experiment}

 The scaling behavior found in the previous section 
can be divided into three distinct regimes (Fig.~6):
\begin{enumerate}
   \item Low--Salt regime $c_b\ll p|y_s|/8\pi l_B a^2$.
   \item First salt--rich (SR I) regime, where  
          $c_b\gg p|y_s|/8\pi l_B a^2$ and $p\ll p^*=(c_b v)^{1/2}$
          (weak polyelectrolytes).
   \item Second salt--rich (SR II) regime, where
          $c_b\gg p|y_s|/8\pi l_B a^2$ and $p\gg p^*$
          (strong polyelectrolytes).
\end{enumerate}

Our scaling results are in good agreement
with  adsorption experiments,
although in experiments the charge distribution of the 
polyelectrolytes can be more complicated. 

\subsection{Low--Salt Regime}

Denoyel et al. \cite{Denoyel} have studied the adsorption
of heteropolymers made of neutral (acrylamide) 
and cationic monomers (derived from chloride acrylate).
The fractional charge was fixed during the polymerization 
process 
and varied from $p=0$  to 1.
Since the salt amount in their experiment was quite low: 
$1.2$mM corresponding to $\kappa_s^{-1}\simeq 90$$\AA$, 
their experimental range satisfies the 
 low--salt conditions.
 Indeed, the measured $\Gamma$ (Table II in ref.~\cite{Denoyel}) 
exhibits a $p^{-1/2}$ dependence as 
in eq.~\ref{scalingGamma1}.

\subsection{Weak Polyelectrolytes: Effect  of Salt}

 Shubin and Linse \cite{Shubin} adsorbed another cationic 
derivative of poly(acrylamide) on silica.
The fractional charge was fixed at a low value ($p=0.034$),
while the salt concentration varied from $c_b=0.1$mM to 
 $c_b\approx 0.2$M.
Ellipsometry was used to measure  $\Gamma$
 and $D$ of the adsorbed layer 
as function of the salt concentration.
This low charge fraction belongs to
the left side (low $p$) of Fig.~6. The experimental 
behavior is similar 
to our predictions as shown in Fig.~5 for weak polyelectrolytes.
At low electrolyte concentration ($c_b<1$mM), 
the adsorbed amount is essentially constant
 and decreases at higher salt concentration (SR I
regime of Fig.~6).
Similar behavior was obtained both by numerical calculations 
using the multi--Stern layer model 
\cite{Shubin,vandeSteegSim,Linse}, and in other
adsorption experiments 
of cationic potato starch \cite{vandeSteegExp}.

\subsection{Strong Polyelectrolytes: Effect of Salt}

   Kawaguchi et al. \cite{Kawaguchi} measured the adsorption of 
a highly charged polyelectrolyte (PVPP) on silica surfaces.
Due to the high ionic strength  this system belongs to
the SR II regime.
Indeed,  $\Gamma\sim\sqrt{c_b}$ was found 
in agreement with our prediction.
   Meadows et al. \cite{Meadows} 
also performed adsorption experiments 
with highly charged
 ($p=0.9$) hydrolyzed 
poly(acrylamide).
The adsorbed amount $\Gamma$ and the width of the 
adsorbed layer $D$  were found to increase upon addition of salt.
Qualitatively, this agrees with our prediction in the SR II regime.
However, 
the measured power laws are weaker than our predictions.
A simple power law fit of their  salt dependence gives 
 $\Gamma\sim c_b^{1/4}$ as compared to our $ c_b^{1/2}$ prediction.
This behavior is intermediate
 between the salt free and SR II regimes.

\subsection{Effect of Charge Fraction}

   Peyser and Ullman \cite{Peyser} studied the adsorption 
of PVP on a glass surface as function of the charge fraction 
for three different salt concentrations.
The system belongs to the right side ($p\alt 1$) of Fig.~5
between the low--salt and SR II regimes.
As expected
$\Gamma$ increases 
with $c_b$ and decreases with 
$p$. Moreover, it is possible to fit
the data to a simple scaling law of the form
$\Gamma\sim c_b^{1/4}/p^{1/2}$. 
Our  scaling results do not fit very
well these experiments  which lie in the intermediate regime,
between the low--salt and SR II regimes.

In experiments on annealed polyelectrolytes 
\cite{Blaakmeer,Hoogeveen}, the polymer charge 
can be tuned by the pH of the solution (eq.~\ref{pHpK}).
The behavior then shifts continuously from the SR I 
to the SR II regimes.
   For example, Blaakmeer et al. \cite{Blaakmeer} used 
polyacrylic acid which is neutral (no dissociation) 
at low pH but becomes 
negatively charged (strong dissociation) at higher pH.
As predicted by eq.~\ref{scalingGamma2} (see also  Fig.~4b), 
a non-monotonous dependence of $\Gamma$ 
on the pH was observed, with a maximum below the pK$_0$.
This effect had been already verified by numerical calculations 
based on the multi--Stern layer model 
\cite{Blaakmeer}.

   A similar maximum in $\Gamma$ was also observed in  
adsorption experiments of proteins  \cite{Norde}
and diblock copolymers 
with 
varying ratios between the charged and neutral blocks 
\cite{Hoogeveen} and may be interpreted using similar considerations.


\section{Conclusions}
\label{conclusions}

 In this work we use simple arguments to derive scaling laws 
describing the adsorption of polyelectrolytes on a single 
charged surface  held at a constant potential. 
We obtain expressions for  the 
amount of adsorbed polymer $\Gamma$ and the width $D$ 
of the adsorbed layer, as a function of the fractional charge $p$ 
and the salt concentration $c_b$. 
   In the low--salt regime a $p^{-1/2}$ dependence  
of $\Gamma$ is found. It is supported by our numerical solutions
of the  profile equations \ref{PBs}, \ref{SCFs} and is   
 in agreement with experiment \cite{Denoyel}.
   This  behavior is due to strong Coulomb repulsion between 
adsorbed monomers in the absence of salt.
As $p$ decreases, the adsorbed amount increases 
until the electrostatic attraction becomes  weaker 
than the excluded volume repulsion,
at which point, $\Gamma$ starts to decrease rapidly.
  At high salt concentrations we obtain two limiting behaviors:
(i) for weak polyelectrolytes, $p\ll p^*= ({c_b v})^{1/2}$, 
the adsorbed amount increases with the fractional charge 
and decreases with the salt concentration, $\Gamma\sim p/\sqrt{c_b}$, 
due to the monomer--surface electrostatic attraction.
(ii) For strong polyelectrolytes, $p\gg p^*$, 
the adsorbed amount decreases with the fractional charge 
and increases with the salt concentration, $\Gamma\sim \sqrt{c_b}/p$, 
due to the dominance of the monomer--monomer electrostatic repulsion.
Between these two regimes  we find that
the adsorbed amount reaches a maximum
in agreement with experiments \cite{Blaakmeer,Hoogeveen}.

   The scaling approach
can serve as a starting point for further 
investigations. Special attention should be directed to the crossover
regime where $D$ and $\kappa_s^{-1}$ are of comparable size.
At present, it is not clear whether the intermediate regime
represents simply a crossover between
regimes or is a 
scaling regime on its own.
Another important question addresses the relative
importance of  attractive versus repulsive 
forces between two charged surfaces in presence of
a polyelectrolyte solution.
Finally,  our approach could be used in  non flat 
geometries 
such as spheres (colloidal particles) and cylinders.

\vspace*{0.5cm}
\noindent
{\em Acknowledgments}

   We would like to thank L. Auvray, M. Cohen-Stuart, 
J. Daillant,
H. Diamant, P. Guenoun, Y. Kantor, P. Linse, 
P. Pincus,  S. Safran and C. Williams for useful discussions.  
   Two of us (IB and DA) would like to thank 
the Service de Physique Th\'eorique (CE-Saclay) 
and one of us (HO) 
the Sackler Institute of Solid State Physics (Tel Aviv University)
for their hospitality.
   Partial support from the German-Israel Foundation (G.I.F) 
under grant No. I-0197 
and the U.S.-Israel Binational Foundation (B.S.F.) 
under grant No. 94-00291 is gratefully acknowledged. 


\section*{Figure Captions}
\pagestyle{empty}

\noindent {\bf Fig.~1}:
Schematic view of a polyelectrolyte solution 
in contact with a flat surface at $x=0$.
The solution contains polyelectrolyte chains and 
small ions. In our model, the surface 
is held at a constant potential.

\bigskip
\noindent {\bf Fig.~2}: 
   Adsorption profiles obtained by numerical solutions of 
eqs.~\ref{PBs},\ref{SCFs}
for several sets of physical parameters in the low--salt limit.
The polymer concentration scaled by its bulk value $\phi_b^2$ 
is plotted as a function of the distance from the surface.
  The different curves correspond to:
 $p=1$, $a=5$$\AA$\ and $y_s=-0.5$ in units of $k_BT/e$ (solid curve);
 $p=0.1$, $a=5$$\AA$\ and $y_s=-0.5$ (dots)
 $p=1$, $a=5$$\AA$\ and $y_s=-1.0$ (short dashes);
 $p=1$, $a=10$$\AA$\ and $y_s=-0.5$ (long dashes);
 and $p=0.1$,$a=5$$\AA$\ and $y_s=1.0$ (dot--dash line).
For all cases  $\phi_b^2=10^{-6}$$\AA$$^{-3}$, $v=50$$\AA$$^3$,
 $\varepsilon=80$,
 $T=300$K and $c_b=0.1$mM.

\bigskip
\noindent {\bf Fig.~3}: 
  Scaling behavior of polyelectrolyte adsorption in the 
low--salt regime (eqs.~\ref{scalingD},\ref{scalingcm}). 
(a) The rescaled electrostatic potential $\psi(x)/|\psi_s|$ 
as a function of the rescaled distance $x/D$.
(b) The rescaled polymer concentration 
 $c(x)/c_m$ as a function of the same rescaled distance.
The profiles are taken from Fig.~2 (with the same notation).
The numerical prefactors of the linear $h(x/D)$ profile
were used in the calculation of $D$ and $c_m$.

\bigskip
\noindent {\bf Fig.~4}: 
   Typical adsorbed amount $\Gamma$ as a function of
(a) the charge fraction $p$ and 
(b) the ${\rm pH -pK}_0$ of the solution 
for three different salt concentrations (eq.~\ref{scalingGamma2}). 
The insets correspond to the low--salt regime 
(eq.~\ref{scalingGamma1}).
The parameters used for $\varepsilon$, $T$ and $v$ 
are the same as in Fig.~2, while $y_s=-0.5$ and $a=5$$\AA$.
The bulk concentration $\phi_b^2$ is assumed to be much smaller 
than $c_m$.
 The numerical prefactors of the linear $h(x/D)$ were used.

\bigskip
\noindent {\bf Fig.~5}: 
    The effect of salt concentration on the adsorption. 
The solid curves correspond to the scaling relations in 
the salt--rich regime (eqs.~\ref{scalingD2},\ref{scalingGamma2}) 
and the dashed curves correspond to a simple 
numerical interpolation between the 
salt--rich and the low--salt regimes.
    (a) Adsorbed amount $\Gamma$ as a function of 
the salt concentration $c_b$ (eq.~\ref{scalingGamma2})
for $p=0.01$ and $0.25$.
    (b) Adsorption length $D$ as a function of 
the salt concentration $c_b$ (eq.~\ref{scalingD2})
for $p=0.1$.
The Debye--H\"uckel screening length $\kappa_s^{-1}$ is 
also plotted (dots). The low--salt (salt--rich) regime
applies when $D \ll \kappa_s^{-1}$ ($D \gg \kappa_s^{-1}$).
The parameters used are:
 $\varepsilon=80$, $T=300$\,K $v=50$$\AA$$^3$, $a=5$$\AA$, $y_s=-2.0$
and the numerical prefactors of the linear $h(x/D)$.

\bigskip
\noindent {\bf Fig.~6}: 
  Schematic diagram of the different adsorption regimes 
as function of the charge fraction $p$ 
and the salt concentration $c_b$.
Three regimes can be distinguished: 
(i) the low--salt regime $D\ll\kappa_s^{-1}$;
(ii) the salt--rich regime (SR I) $D\gg\kappa_s^{-1}$
for weak polyelectrolytes $p\ll p^*=({c_b v})^{1/2}$;
and (iii) the salt--rich regime (SR II) $D\gg\kappa_s^{-1}$
for strong polyelectrolytes $p\gg p^*$.

\pagebreak \vfill

\begin{figure}[tbh]
  \vspace{12\baselineskip}
  {\Large Fig.~1} 
  \bigskip\bigskip\bigskip

  \epsfxsize=0.5\linewidth
  \centerline{\hbox{ \epsffile{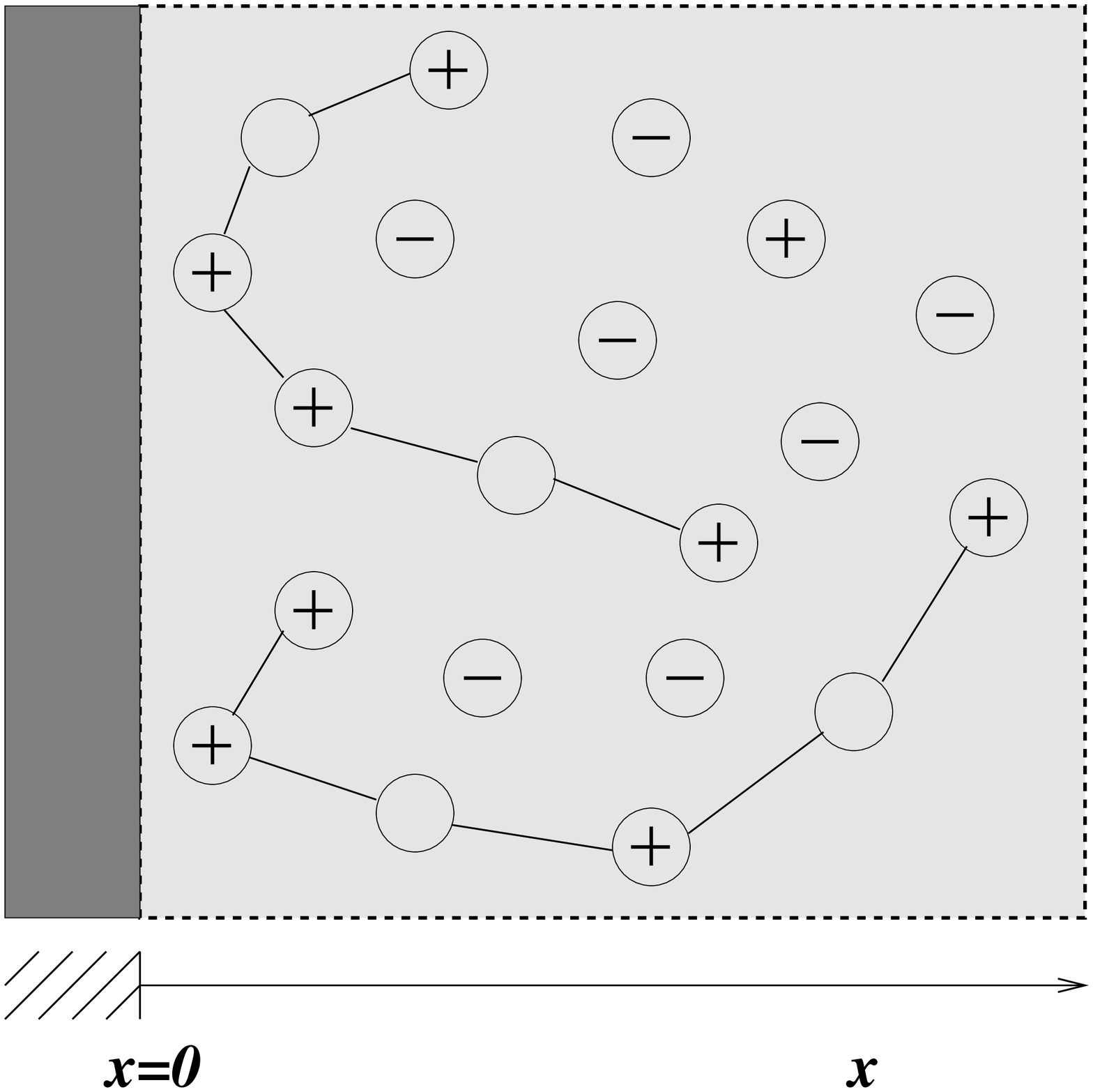} }}
\end{figure}
\vfill


\begin{figure}[tbh] 
  {\Large Fig.~2} 
  \bigskip\bigskip\bigskip

  \epsfxsize=0.5\linewidth
  \centerline{\hbox{ \epsffile{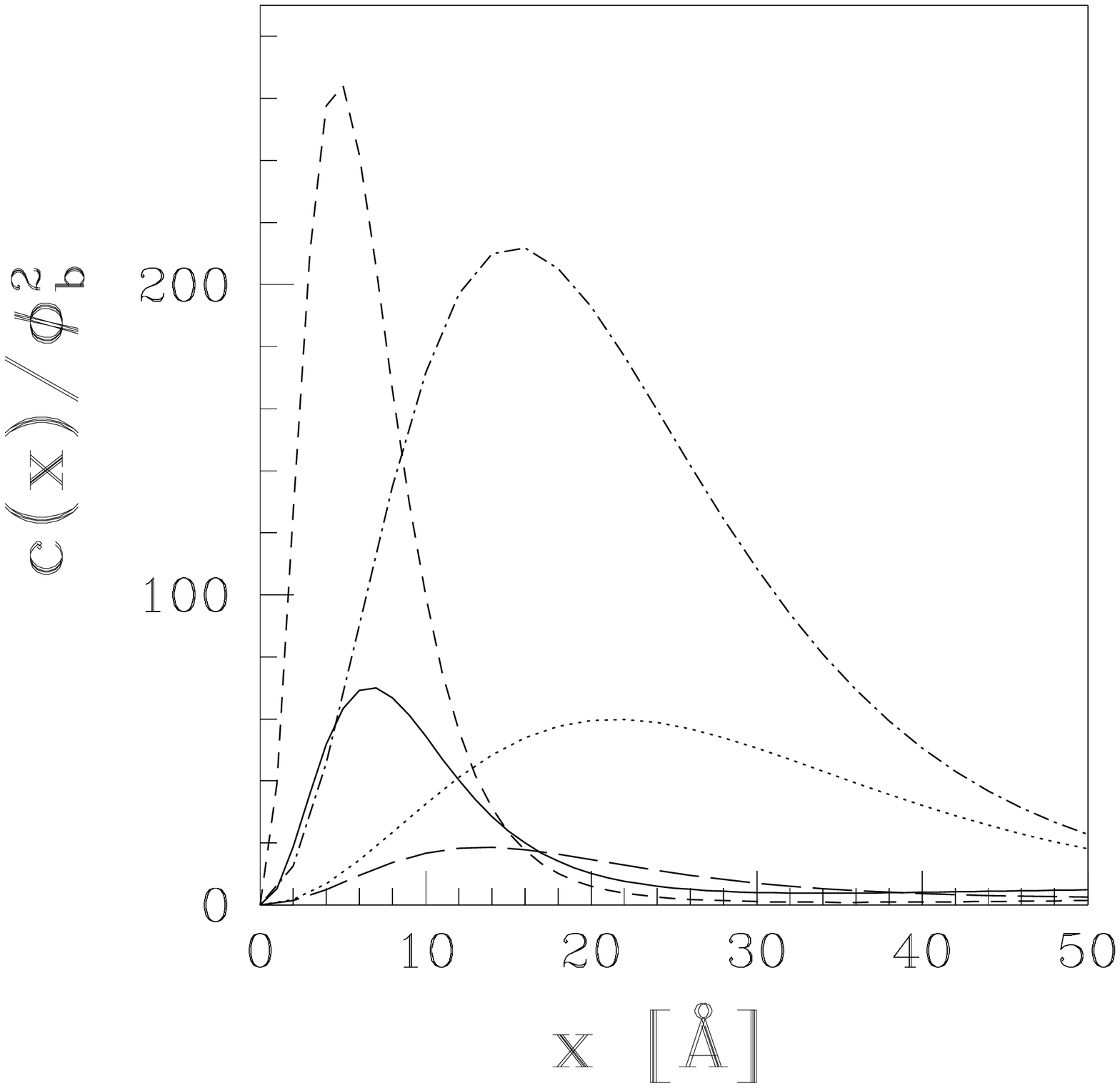} }}
\end{figure}
\vfill

\pagebreak \vfill

\begin{figure}[tbh]
  {\Large Fig.~3a} 
  \bigskip\bigskip\bigskip

  \epsfxsize=0.5\linewidth
  \centerline{\hbox{ \epsffile{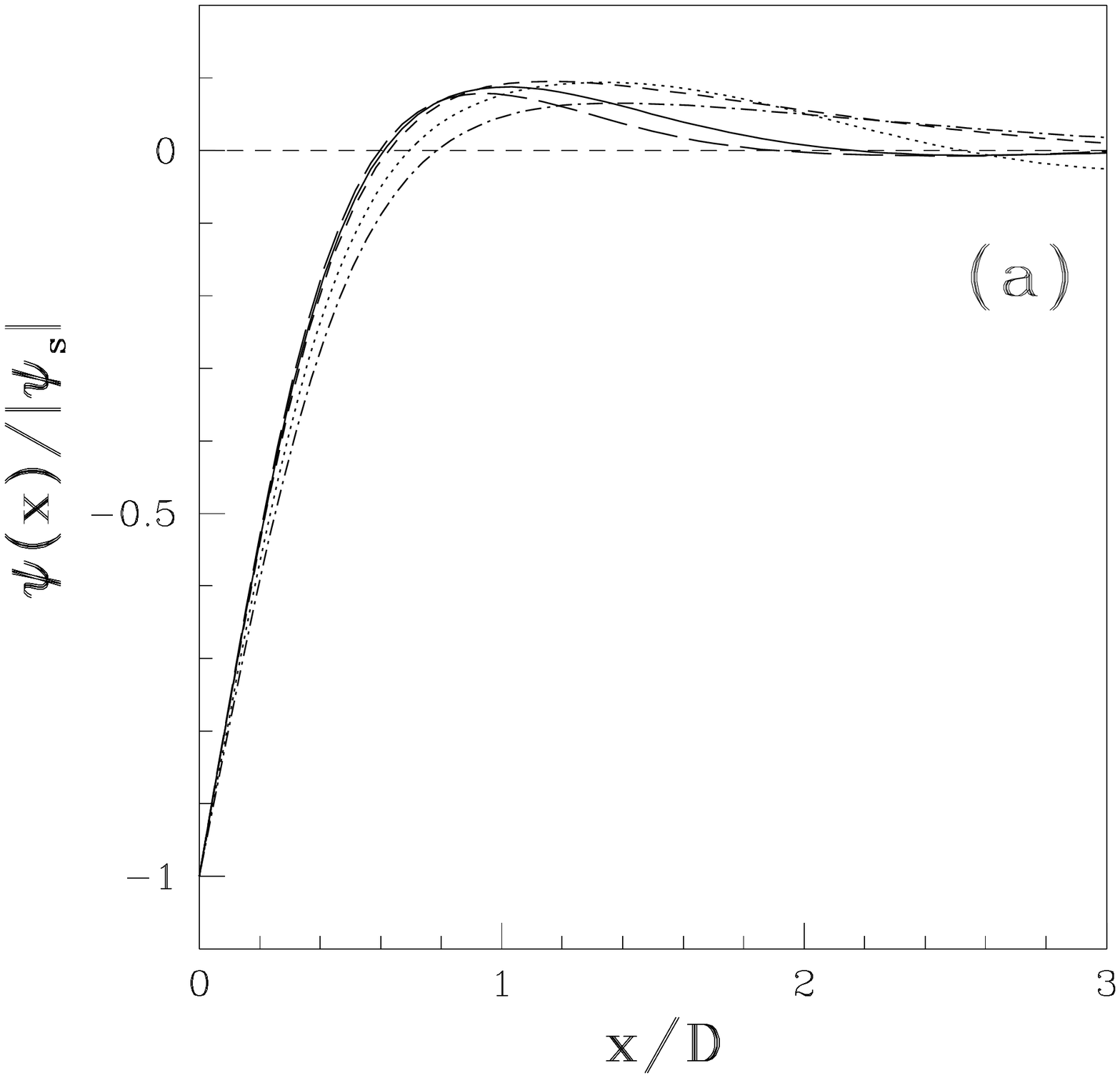} }}
\end{figure}
\vfill

\begin{figure}[tbh]
  {\Large Fig.~3b} 
  \bigskip\bigskip\bigskip

  \epsfxsize=0.5\linewidth
  \centerline{\hbox{ \epsffile{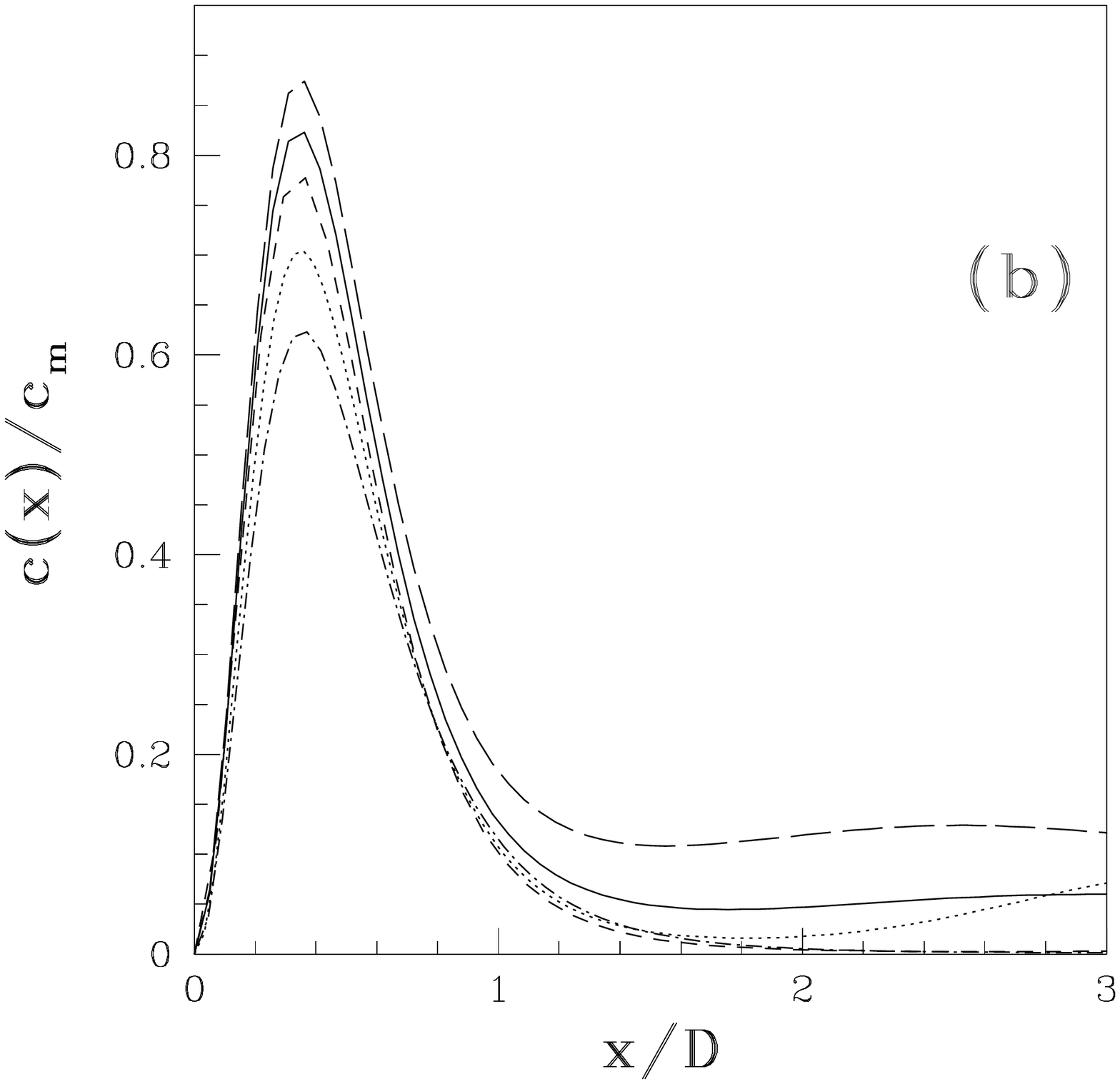} }}
\end{figure}
\vfill

\pagebreak \vfill

\begin{figure}[tbh]
  {\Large Fig.~4a} 
  \bigskip\bigskip\bigskip

  \epsfxsize=0.5\linewidth
  \centerline{\hbox{ \epsffile{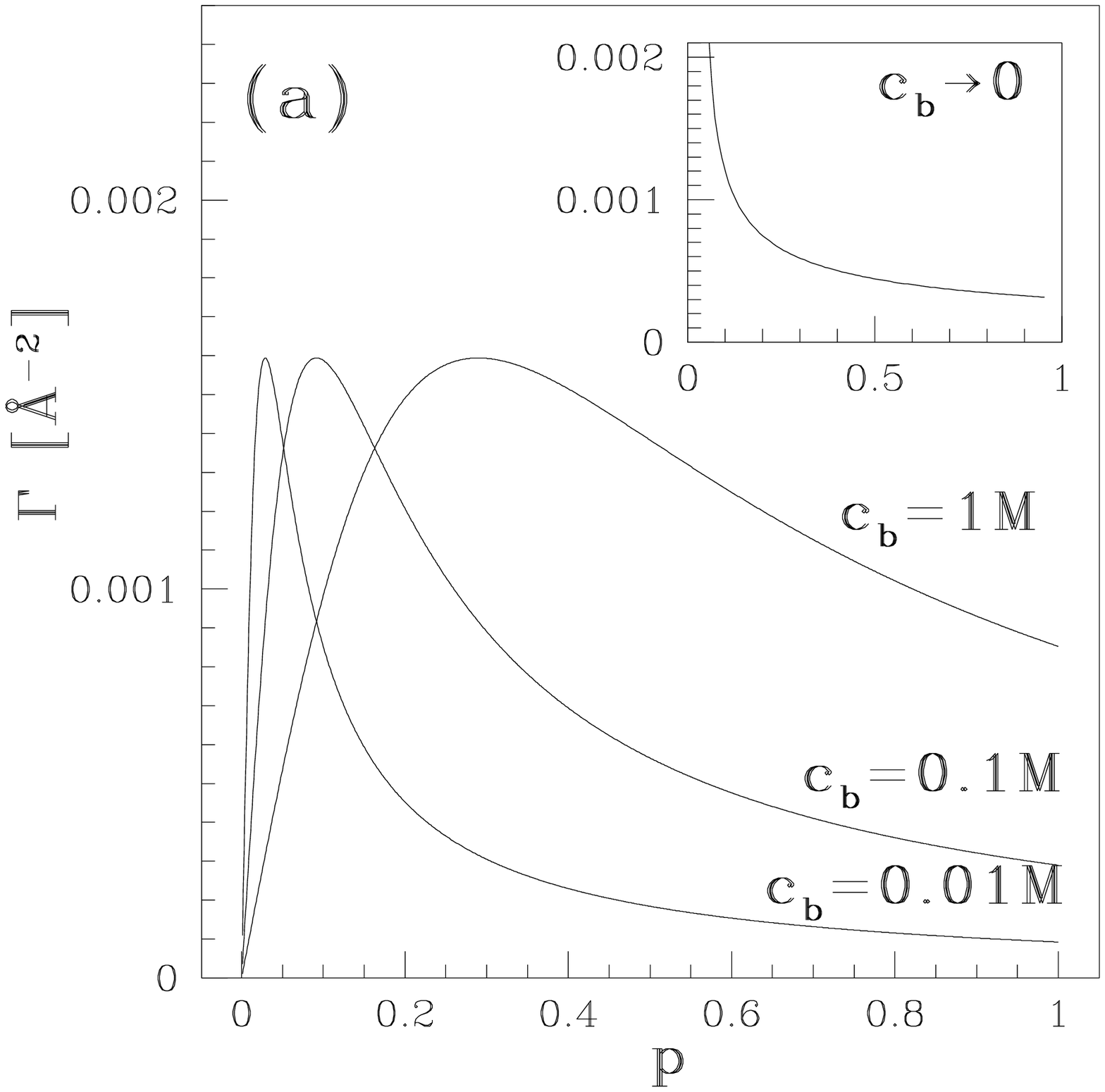} }}
\end{figure}
\vfill

\begin{figure}[tbh]
  {\Large Fig.~4b} 
  \bigskip\bigskip\bigskip

  \epsfxsize=0.5\linewidth
  \centerline{\hbox{ \epsffile{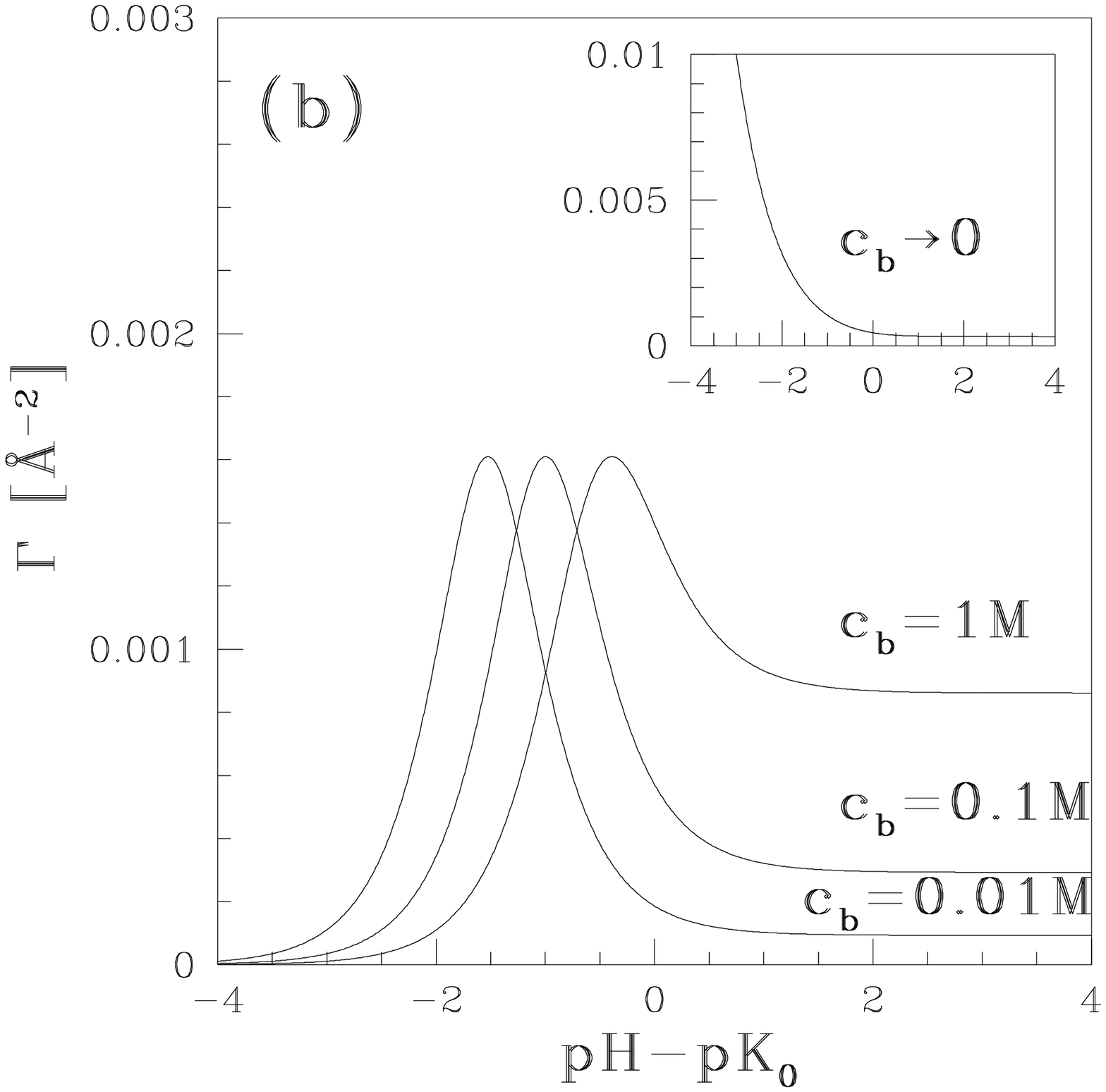} }}
\end{figure}
\vfill

\pagebreak \vfill

\begin{figure}[tbh]
  {\Large Fig.~5a} 
  \bigskip\bigskip\bigskip

  \epsfxsize=0.5\linewidth
  \centerline{\hbox{ \epsffile{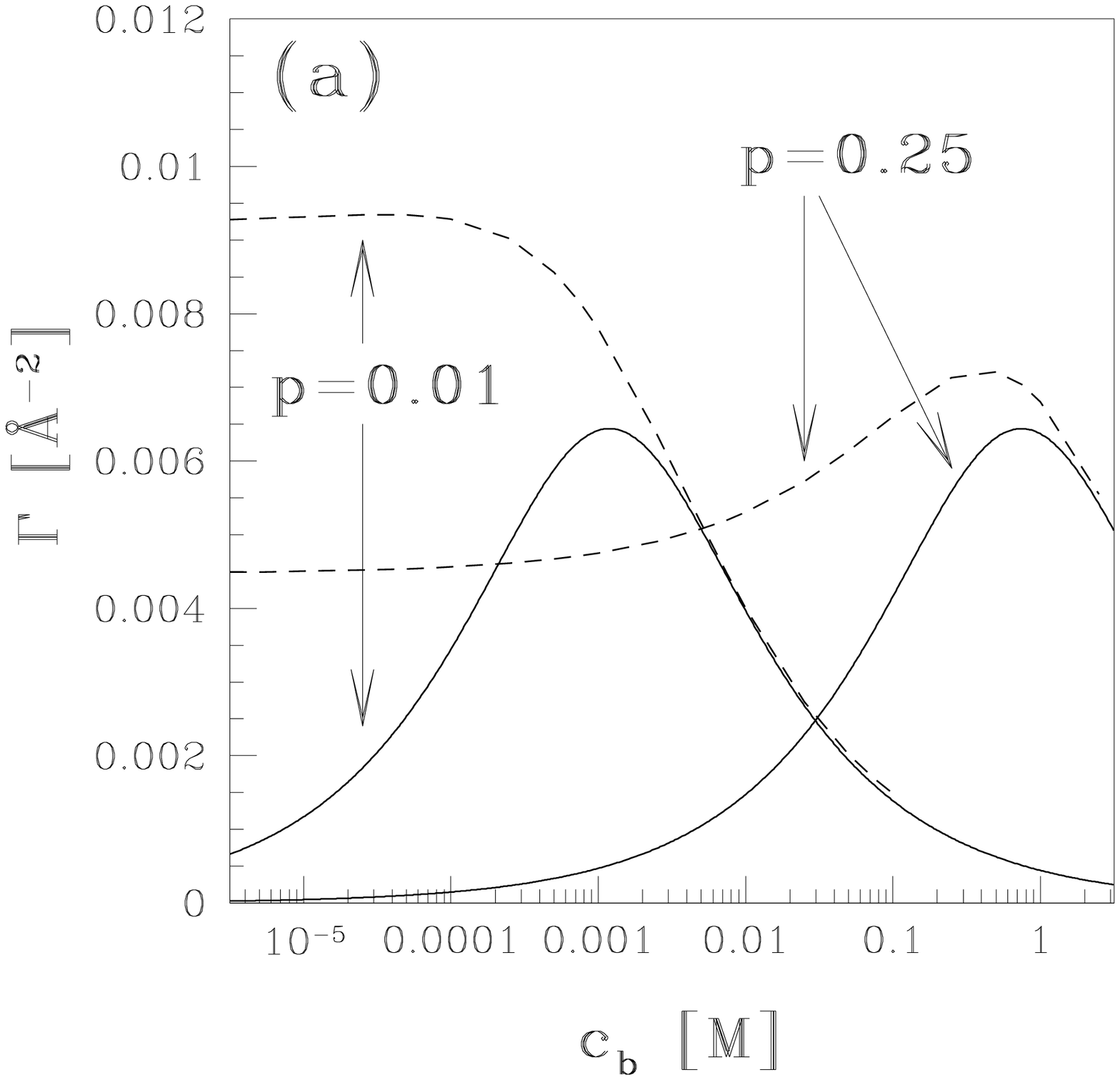} }}
\end{figure}
\vfill

\begin{figure}[tbh]
  {\Large Fig.~5b} 
  \bigskip\bigskip\bigskip

  \epsfxsize=0.5\linewidth
  \centerline{\hbox{ \epsffile{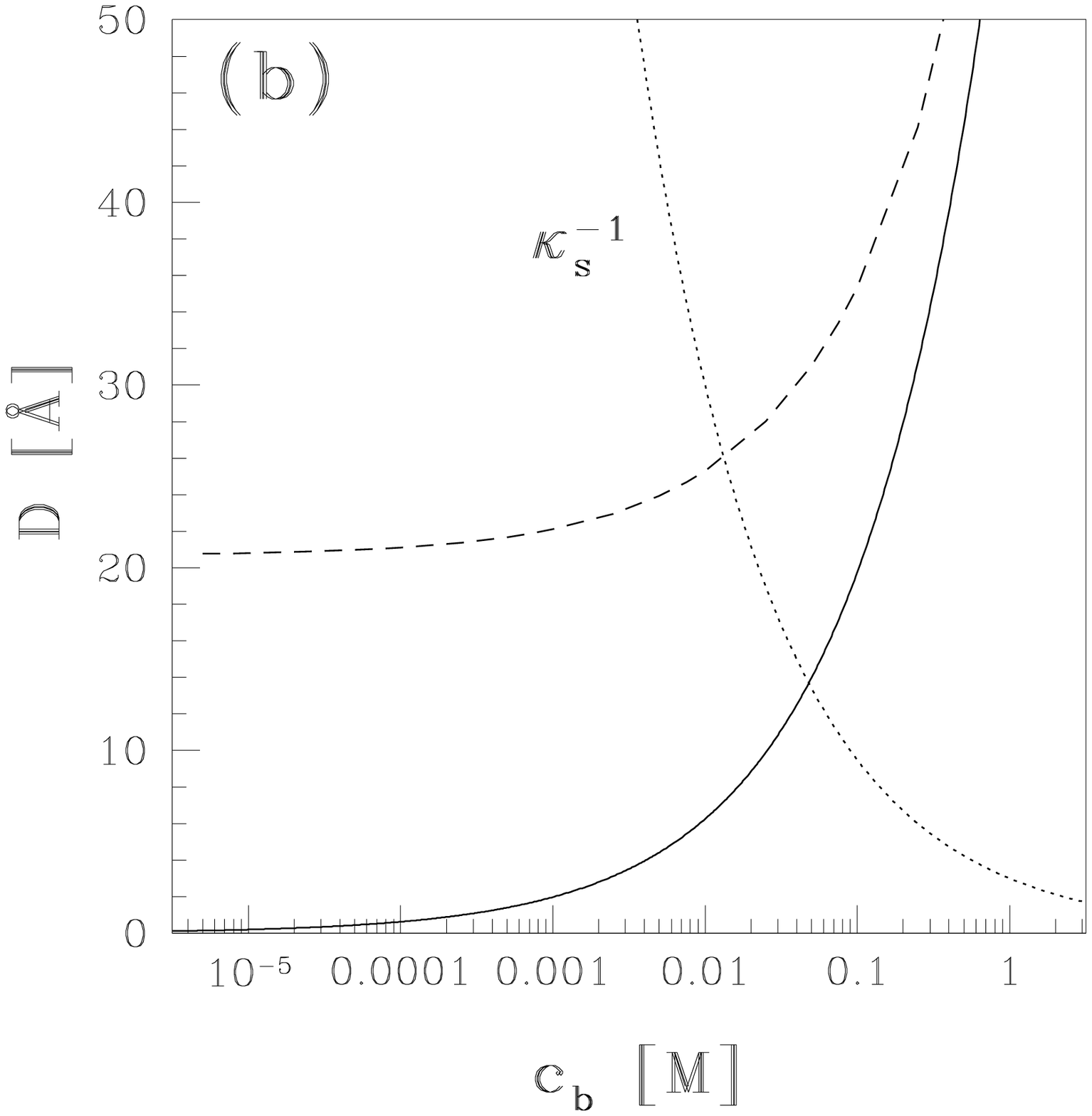} }}
\end{figure}
\vfill

\pagebreak \vfill

\begin{figure}[tbh]
  {\Large Fig.~6} 
  \bigskip\bigskip\bigskip

  \epsfxsize=0.5\linewidth
  \centerline{\hbox{ \epsffile{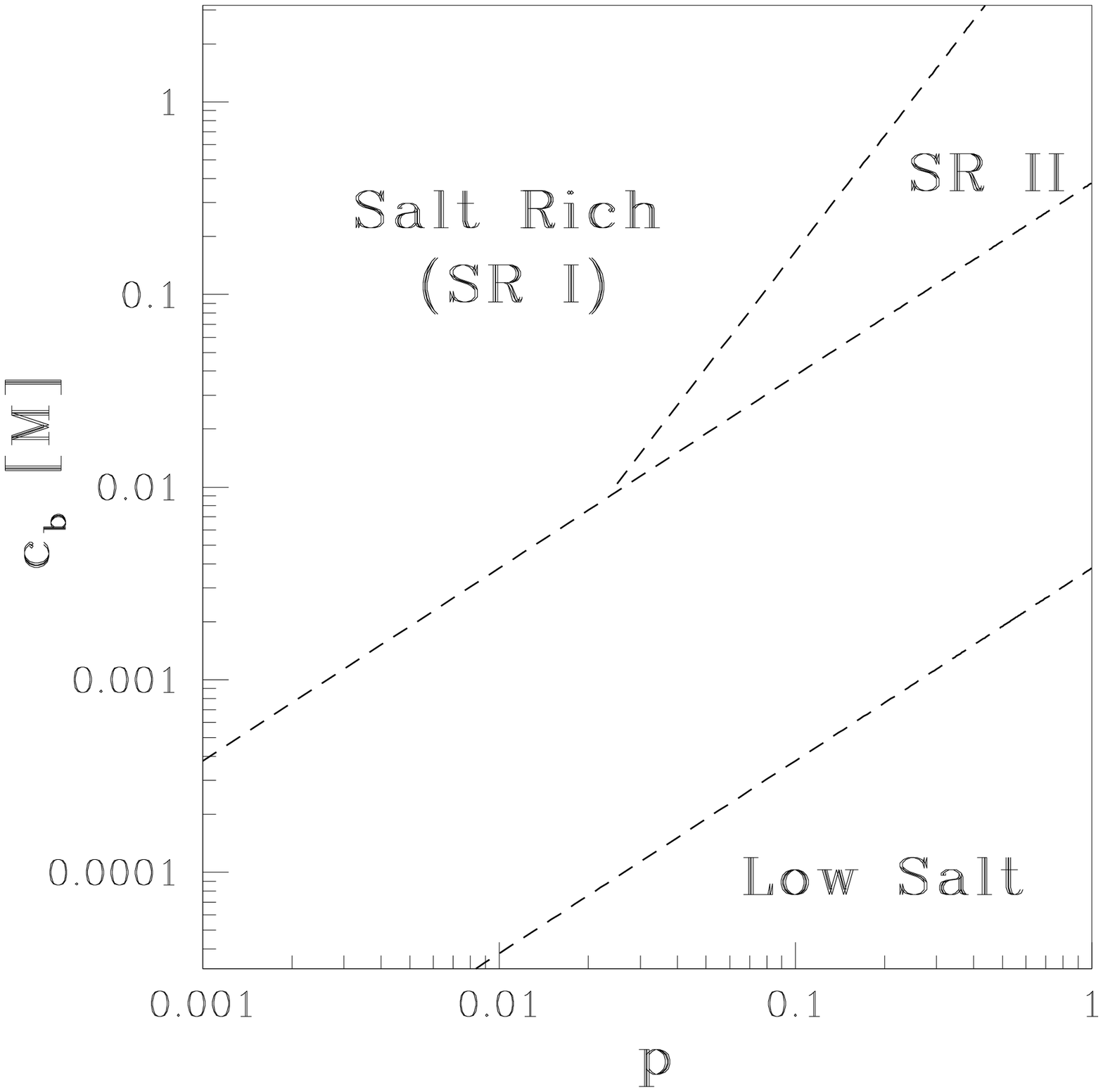} }}
\end{figure}
\vfill


\begin{thebibliography}{99}
\bibitem{elec_bc}
 It is also  possible to use
 a constant surface charge density 
 $\sigma$. The boundary condition then becomes
 $ \left. \psi' \right|_s = - {4\pi \sigma}/{\varepsilon} $.
 In real systems, some of the ionized surface sites 
can be neutralized by the  binding of small ions from the solution.
The actual surface charge is not fixed but 
depends on the surface potential \cite{IsraelachviliBook}.
Neither the surface potential nor the 
surface charge are fixed, and the boundary conditions are mixed.
The choice of simple boundary conditions is only an approximation
of a real surface--solution system. The quality of the approximation
depends on the details of the experimental system.
\bibitem{coefficient}
For a given profile 
$h(x/D)$, the electrostatic potential 
at a distance $x$ from the surface can be calculated
by integrating the charge density from the surface up to the
distance $x$.
%
The surface charge
density is determined from charge neutrality which is imposed  by setting 
the electric field to zero at $x=D$. The electrostatic term in the free
energy $f_{el}$ is then obtained by integrating $pec_m h^2(x/D)\psi(x)$.
\bibitem{interpolation}
    In order to interpolate between the low--salt 
and the salt--rich regimes, we replace $\kappa_s^{-1}$ in
the free energy, eq. \ref{scalingF2} with 
$D_{\rm eff} \equiv 1/(1/D+\kappa_s)$, and minimize the resulting 
free energy numerically.
\end{thebibliography}
\end{document}